\documentclass[reqno]{article}

\usepackage{amsfonts,amssymb,amsmath,amsthm,epsfig,cite}
\usepackage{xcolor}

\newcommand{\oB}{\vert_{\partial\mathcal{M}}}

\newcommand{\boundary}{\partial \mathcal M}
\newcommand{\uB}{\boundary^U}
\newcommand{\lB}{\boundary^L}
\DeclareMathOperator{\extdm}{d}
\newcommand{\extd}{\extdm \!}
\def\cM{{\cal M}}

\title{\vspace{-8ex} \hfill {\normalsize MPP-2007-171}\\[6ex]Two-Dimensional Quantum Gravity with Boundary}

\bigskip
\author{
Luzi Bergamin \\
\small\it ESA Advanced Concepts Team, ESTEC -- DG-PI,\\[-1.mm]
\small\it Keplerlaan 1, 2201 AZ Noordwijk, The
Netherlands,  \\[-1.mm]
\small\it email: Luzi.Bergamin@esa.int\\
Ren\'{e} Meyer\\[-1.mm]
\small\it Max Planck Institut f\"{u}r Physik,\\[-1.mm]
\small\it F\"{o}hringer Ring 6, 80805 \small\it M\"{u}nchen, Germany,\\[-1.mm]
\small\it email: meyer@mppmu.mpg.de}
\date{ }

\begin{document}

\maketitle
\bigskip

\begin{abstract}

Using the recently found first order formulation of two-dimensional dilaton gravity with boundary, we perform a Hamiltonian analysis and subsequent path integral quantization. The importance of the boundary terms to obtain the correct quantum result are outlined and the quantum triviality of the theory is shown to hold with this modification as well. We compare with recent classical results and comment on further applications.

\end{abstract}

\section{Introduction}
Two-dimensional dilaton gravity \cite{Grumiller:2002nm,Grumiller:2006rc} with boundaries recently attracted some attention. On the one hand, an eventual boundary was identified with the black hole horizon to attack the problem of black hole entropy \cite{Carlip:2004mn,Bergamin:2005pg,Bergamin:2006zy,Carlip:2006fm}, on the other hand the boundary was used as regulator to ensure a correct semi-classical expansion and played an important role in the calculation of thermodynamical properties \cite{Davis:2004xi,Grumiller:2007ju,Bergamin:2007sm}. Both applications naturally raise the question how such a system could be quantized in a systematic way. For the simpler case without boundary, a first order formulation combined with a temporal gauge has proved to be extremely powerful \cite{Kummer:1997hy,Kummer:1998zs,Grumiller:2002nm} in this respect. In this paper we make a first step to extend this analysis to spacetimes with boundary.

\section{Two-dimensional dilaton gravity with boundary}
The action of generalized two-dimensional dilaton gravity in first order formulation reads \cite{Grumiller:2002nm}
\begin{equation}
 \label{act1}
\mathcal S = - \int_{\mathcal M} \left[ X^a D e_a + Xd\omega + \epsilon\left(U(X) X^+X^-+V(X)\right) \right]\ ,
\end{equation}
where the dilaton $X$ and the Lagrange multipliers for torsion $X^a$ are scalar fields, $\omega$ and $e_a$ are the \emph{independent} spin connection and the dyad, resp., the
latter yields the volume form $\epsilon = \epsilon^{ab} e_b \wedge e_a$. Though this action does not have any physical degrees of freedom in the bulk it is not completely trivial, as there always exists a conserved quantity (Casimir function)
\begin{equation}
\label{casimir}
 \mathcal C = e^{Q} X^+ X^- + w\,,\quad \text{with}\quad  Q(X)= \int\limits^X \extd y U(y)\ , \  w(X)= \int\limits^X \extd y e^{Q(y)} V(y)\ .
\end{equation}
This conserved quantity encodes global nontrivial information (essentially the black hole mass) and classically can be chosen freely. For further conventions and definitions we refer to \cite{Grumiller:2002nm}.

Once spacetime is considered with boundaries, a consistent solution of the variational problem has to be found. To this end we add a boundary term $\int_{\partial \mathcal M} X\omega_{\|}$, which moves the exterior derivative in the second term of \eqref{act1} onto the dilaton $X$. Taking the variation with respect to $X$ and $e_a$, the conditions $\omega \delta X\oB = 0$, $X^a \delta e_a \oB = 0$ arise in order for all variations to be well defined. Here, they are solved by imposing Dirichlet boundary conditions on the dilaton and the dyad.

Though this allows a consistent variation of the action, additional boundary terms can be added that depend exclusively on the quantities held fixed at the boundary. One such term is needed to restore local Lorentz invariance along the boundary and to ensure the equivalence of the first order formulation with the standard Gibbons-Hawking prescription \cite{Bergamin:2005pg}, which yields the action
\begin{equation}
\label{act2}
  \mathcal S  = \mathcal S_{\mbox{\tiny bulk}} + \mathcal S_{\mbox{\tiny boundary}} = - \int_{\mathcal M} \left[ X^a D e_a + Xd\omega + \epsilon(UX^+X^-+V) \right]  + \int_{\partial \mathcal M} \left[ X\omega_{\|} + \frac{1}{2} Xd\ln \frac{e_{\|}^{+}}{e_{\|}^{-}}\right]\ .
\end{equation}
A second term is a Hamilton-Jacobi counterterm and is necessary to ensure a consistent semi-classical expansion \cite{Davis:2004xi,Grumiller:2007ju,Bergamin:2007sm}. In its derivation it has been assumed that the dilaton is constant along the boundary, which is sufficient for all practical purposes. Still, the corresponding boundary action
\begin{equation}
 \mathcal S_{\mbox{\tiny boundary}} = \int_{\partial \mathcal M} \left[ X\omega_{\|} +  \frac{1}{2} Xd\ln \frac{e_{\|}^{+}}{e_{\|}^{-}} - 2 \sqrt{e_{\|}^{+}e_{\|}^{-}} \sqrt{-w(X)e^{-Q}} \right]\ ,
\end{equation}
contains an additional assumption which however is not relevant to the current discussion.

\section{Path-integral quantization of the action}
In \cite{Bergamin:2005pg} a Hamiltonian analysis of the action \eqref{act2} has been performed, whereby the boundary was considered either as a horizon or as a timelike surface. There arose considerable technical complications compared to the case without boundary: the boundary conditions have to be implemented by new boundary constraints, which turn some of the bulk constraints from first into second class; at the same time a Dirac procedure remains problematic, as the ``Dirac matrix'' has support at the boundary, only. This did not affect the results in \cite{Bergamin:2005pg}, as the authors were interested in the classical, reduced phase-space. However, in a BRST quantization this complication becomes relevant.

To enable a quantization along the lines of \cite{Kummer:1997hy,Kummer:1998zs,Grumiller:2002nm} we propose a procedure in the spirit of \cite{Carlip:2004mn} rather than \cite{Bergamin:2005pg}. There the boundary was not considered as timelike, but rather spacelike, where ``time''  $x^0$ refers to the evolution parameter of the Hamiltonian analysis. From now on we will assume that the boundary is at a fixed value of $x^0$. The boundary conditions, now being initial or final conditions thereof, no longer affect the constraint algebra and thus in the bulk all known results can be taken over. Our boundary has an upper part $\partial \cM^U$ at time $x^0_U$, which comes with a sign change in the action as our boundary is oriented inwards, and a lower one, $\partial \cM^L$ at time $x^0_L$.\footnote{For pedagogical reasons we consider both, an upper and lower boundary, as only in this way the different roles of them in the full quantum result become transparent. Of course, the boundary data can be chosen only on one of them, while at the other boundary it is then defined uniquely via the equations of motion.}
 The action \eqref{act2} in components then reads
\begin{multline}
\label{act3}
 \mathcal S =-\int\limits_{\mathcal{M}}  \extd^2x \left[ X_a (D_\mu e_\nu^a) \tilde\epsilon^{\mu\nu}
+X\partial_\mu \omega_\nu \tilde \epsilon^{\mu\nu}
+ e(U(X)X^+X^- + V(X)) \right] \\
+ \int_{\uB \cup \lB} \extd x^1\left[X \omega_1 + \frac 12 X \partial_1 \ln \frac{e_1^+}{e_1^-} - 2 \sqrt{-e_1^+ e_1^- e^{-Q} w} \right]\ ,
\end{multline}
Following the standard quantization procedure \cite{Grumiller:2002nm} we choose coordinates and momenta as
\begin{align}
 q_i &= (\omega_1, e_1^-,e_1^+)\ , & \bar{q}_i &=(\omega_0, e_0^-, e_0^+)\ , \\
 p_i &= (X, X^+,X^-)\ , & \bar{p}_i&\approx 0\ ,
\end{align}
where the $\bar{p}_i$ are primary first-class constraints. The bulk part of the canonical Hamiltonian is the sum over three secondary first-class constraints associated with local Lorentz and diffeomorphism invariance,
\begin{equation}
 \mathcal H = - \bar q_i G_i \\ .
\end{equation}
Only at the boundaries the additional surface terms from \eqref{act3} have to be added to the Hamiltonian, which however do not influence the BRST construction in any way. We thus can adopt all relevant steps of quantization from the existing results without boundary \cite{Kummer:1997hy,Kummer:1998zs,Grumiller:2002nm}: From the classical Poisson brackets the quantum version is obtained along the standard procedure of Batalin, Vilkovisky and Fradkin \cite{Fradkin:1975cq,Batalin:1977pb,Fradkin:1978xi}. Following \cite{Kummer:1997hy,Kummer:1998zs,Grumiller:2002nm} we employ a multiplier gauge 
\begin{align}
\label{gauge}
 \omega_0 &= 0\ , & e_0^+ &= 0\ , & e_0^- &= 1\ ,
\end{align}
which corresponds to Eddington-Finkelstein gauge for the metric. The ensuing gauge fixed Hamiltonian is used in a path integral
\begin{equation}
 \label{path1}
W[J_i,j_i] = \int \mathcal D(q_i,\bar q_i, p_i, \bar p_i, c_i, p_i^{c}, b_i, p_i^{b}) \exp\left[i \mathcal S^{(1)}\right]\ ,
\end{equation}
where $(c_i, p_i^{c}, b_i, p_i^{b})$ are the ghosts and ghost momenta, resp., and sources $j(x)$ for $q_i$ and $J(x)$ for $p_i$ have been introduced. It is important to realize how the choice of boundary conditions affects the above expression: The path integral measures $D p_1$, $D q_2$ and $Dq_3$ are restricted to contain only paths matching with Dirichlet boundary conditions imposed on these fields.

All ghosts and ghost momenta as well as $\bar q_i$ and $\bar p_i$ can be integrated trivially yielding the simpler path integral
\begin{equation}
 W[J_i,j_i] = \int \mathcal D(q_i,p_i) \det M \exp\left[i \mathcal S^{(2)}\right]\ ,
\end{equation}
where $\det M$ is a functional determinant from integration over the ghosts and
\begin{multline}
\label{path2}
 \mathcal S^{(2)} = \int_{\mathcal M} \extd^2x \left(p_i \dot q_i + q_1 p_2 - q_3 (U p_2 p_3 + V) + j_i q_i + J_i p_i \right)\\
+ \int_{\uB\cup \lB}\extd x^1 \left[p_1 q_1 + \frac 12 p_1 \partial_1 \ln \frac{q_3}{q_2} - 2 \sqrt{-q_2 q_3 e^{-Q} w }\right]\ .
\end{multline}
In the nonperturbative path integral quantization of dilaton gravity in two dimensions \cite{Grumiller:2002nm}, it was crucial that $S^{(2)}$ was linear in the $q^i$ to be able to perform the path integral over these variables. It seems that the boundary terms spoil this feature as they depend nonlinearly on $q_2$ and $q_3$. This is, however, not the case since in these potentially dangerous terms the variables appearing nonlinearly are those, which are fixed at the boundary and therefore are not path-integrated over, as far as the quantum theory in the bulk is concerned.\footnote{Remember that a path integral is only complete by stating the boundary conditions for its fields. In a second step one might try to set up a quantum theory of possible boundary degrees of freedom by integrating over the boundary values of the fields.} So as in \cite{Grumiller:2002nm} we perform the integration over the $q_i$ first, but to do this we have to integrate the kinetic term $p_i \dot q_i$ by parts, which changes the boundary action, thereby becoming independent of $q_1$. This is a nice side-effect as we do not impose Dirichlet boundary conditions on this variable.

In the following we make the simplifying assumption that all sources vanish at the boundary, which is sufficient for most applications. Therefore, the subsequent results strictly apply in the bulk, only. However, as the boundary values of all fields must be compatible with the bulk solution, an analytic continuation of the result in the bulk still allows to derive the correct values at the boundary (cf.\ \cite{Bergamin:2005pg}). Path-integrating of the linearly appearing $q^i$'s yields functional $\delta$-functions in the path integral, enforcing
\begin{align}
\label{solp1}
 p_1 &= B_1 = \bar p_1 + \partial_0^{-1} A_1\ ,\\
\label{solp2}
 p_2 &= B_2 = \bar p_2 + \partial_0^{-1} A_2\ , \\
\label{solp3}
 p_3 &= B_3 = e^{-Q}\left[\partial_0^{-1} A_3 + \bar p_3\right]\ ,
\end{align}
with
\begin{align}
 A_1 &= B_2 + j_1\ , \\
 A_2 &= j_2\ , \\
 A_3 &= e^Q (j_3-V)\ , \\
\label{Qdef}
 Q&= \partial_0^{-1} \left[U(B_1) B_2\right]\,
\end{align}
and $\partial_0 \bar p_i = 0$, but $\partial_1 \bar p_i \neq 0$ in general. 
The Green's function $\partial_0^{-1}$ can be represented as an integral,
\begin{equation}
\label{green}
 (\partial_0^{-1})_{xz} f(z) = \int_{x^0_L}^{x^0} \extd z^0 f(z^0,x^1)\,.
\end{equation}
For convenience the lower integration limit has been chosen as $x^0_L$, as it simplifies the calculations, even though the final result does not depend on this choice. Furthermore it is important to notice that the ambiguity due to the integration constant in \eqref{green} has already been taken care of by the $\bar p_i$. The final generating functional then becomes
\begin{equation}
\label{path3}
 W[J_i,j_i] =  \exp\left[i \mathcal S_{\mbox{\tiny QM}}\right]\ ,
\end{equation}
with the full quantum mechanical action
\begin{multline}\label{eq:effact}
  \mathcal S_{\mbox{\tiny QM}} = \int\limits_{\mathcal M} \extd^2x \left[ B_i J_i + \bar g_i A_i \right]
- \int_{\uB\cup \lB}\extd x^1 \left[B_2 q_2 + B_3 q_3 - \frac 12 p_1 \partial_1 \ln \frac{q_3}{q_2} + 2 \sqrt{-q_2 q_3 e^{-Q} w }\right]\ ,
\end{multline}
whereby in the boundary actions the fixed boundary values for the $q_1$, $q_2$ and $p_1$ are understood while for $B_2$ and $B_3$ the full quantum result from eqs.\ \eqref{solp2} and \eqref{solp3} has to be used. The coefficients of the ``ambiguous terms'' $\bar g_i$ again obey $\partial_0 \bar g_i = 0$, but a dependence on $x^1$ cannot be excluded from the beginning.

\section{Local quantum triviality and equations of motion}
As the generating functional could be evaluated completely in \eqref{path3} it is as well possible to calculate the quantum effective action exactly. This already has been used in \cite{Kummer:1997hy,Grumiller:2002nm,Bergamin:2004us} to prove local quantum triviality of two-dimensional dilaton gravity without matter. Notice that in the definition of the quantum effective action
\begin{equation}
 \Gamma = \mathcal S_{\mbox{\tiny QM}} - \int \extd^2 x \left(j_i \langle q_i \rangle + J_i \langle p_i \rangle \right)
\end{equation}
the ambiguous terms contribute as surface terms, only. Expressing all sources $j_i$ in terms of the mean fields by means of \eqref{solp1}-\eqref{solp3} one finds complete agreement of the bulk action with the classical, gauge fixed action. At the boundary a naive application of our results yields an additional term of the form
\begin{equation}
\label{newboundary}
 \int_{\uB}\extd x^1 \left(\bar g_1 p_1 + \bar g_2 p_2 + \bar g_3 e^Q p_3\right)\ .
\end{equation}
However, we stress again that we cannot expect to reproduce the correct boundary terms at this stage, as we assumed the sources to vanish at the boundary.

\subsection{Expectation values}
Though an agreement of the bulk effective action with the classical action could be derived, the solutions for $\langle q_i(x^0,x^1) \rangle$ and $\langle p_i(x^0,x^1) \rangle$ are yet more general than the classical phase space as the six integrating functions $\bar p_i(x^1)$ and $\bar g_i(x^1)$ have not yet been fixed. A full quantum treatment to fix these constants appropriately is postponed to a future publication, here we impose a simpler, semi-classical technique. Indeed, as the bulk effective action is equivalent to the classical action, the explicit expressions of the mean fields for vanishing sources must be compatible with the classical equations of motion. To derive the ensuing restrictions on the integration functions we derive the exact expressions for all mean fields first.
 
The boundary action is independent of $J_i$ and thus the $\langle p_i \rangle$ are just the known results from the bulk theory \cite{Grumiller:2002nm},
\begin{align}
\label{p1}
 \langle p_1 \rangle &= B_1[j=0] = \bar p_1 + \bar p_2 (x^0-x^0_L) \\
\label{p2}
 \langle p_2 \rangle &= B_2[j=0] = \bar p_2\\
\label{p3}
 \langle p_3 \rangle &= B_2[j=0] = \frac{e^{-Q(x^0)}}{\bar p_2}\left( \bar p_2 \bar p_3 - \left(w(x^0) - w(x^0_L)\right) \right)
\end{align}
In \eqref{p3} we have set both lower integration limits appearing in the definitions in eq.\ \eqref{casimir} to $x^0_L$ for convenience. The associated freedom to scale the conformal factor $\exp(Q)$ by a constant is fully captured in the other integration constants already (cf.\ e.g.\ \cite{Bergamin:2005pg}).

Less straightforward are the $\langle q_i \rangle$. As far as the bulk part of the action is concerned, we can set $J_i=0$ and thus just the ``ambiguous terms'' $\bar{g}_i A_i$ contribute. In the boundary part of \eqref{eq:effact} all contributions may be dropped that depend exclusively on quantities held fixed there. This reduces the relevant part of \eqref{eq:effact} to
\begin{equation}
   \mathcal S_{\mbox{\tiny QM}} = \int_{\mathcal M} \extd^2x \bar g_i A_i + \int_{\uB}\extd x^1 \left[B_2 q_2 + B_3 q_3 \right] -  \int_{\lB}\extd x^1 \left[B_2 q_2 + B_3 q_3\right]\ .
\end{equation}
From \eqref{solp2} and \eqref{solp3} together with the prescription \eqref{green} it is seen that no contribution from the lower boundary can arise to the variation with respect to any point in the bulk. Therefore we may drop this part as well. Furthermore, plugging \eqref{green} into \eqref{solp2} at the upper boundary, the resulting contribution can be absorbed by a redefinition of the coefficient of the ambiguous term $\propto \bar g_2$:
\begin{equation}
\label{g2tilde}
 \tilde g_2 = \bar g_2 + q_2(x^0_U)\ .
\end{equation}
No simple transformation of this kind exists for the remaining boundary contribution $B_3 q_3$, as the factor $e^{-Q}$ in \eqref{solp3} depends on $j_1$ and $j_2$ (cf.\ eq.\ \eqref{Qdef}). After rewriting some boundary terms as integrations over the whole bulk, the final action relevant for the derivation of the $\langle q_i \rangle$ reads
\begin{equation}
  \mathcal S_{\mbox{\tiny QM}} = \int \extd x^1 \left\{ \int_{x^0_L}^{x^0_U} \extd x^0 \left\{ \bar g_1 A_1 + \tilde g_2 A_2 + \left( \bar g_3 + \left[q_3 e^{-Q}\right](x^0_U) \right) A_3 \right\} + \left[q_3 e^{-Q}\right](x^0_U) \bar p_3 \right\}\ .
\end{equation}
Obviously, all variations acting on the sources in $A_3$ in the boundary term are captured by a redefinition of $\bar g_3$ as
\begin{equation}
\label{g3tilde1}
 \tilde g_3 = \bar g_3 + [q_3 e^{-Q}](x^0_U)\ .
\end{equation}
The full expressions for the expectation values now follow from variation w.r.t to the sources $j_i$. It turns out to be useful to define the shifted coefficient $\tilde g_1$ as
\begin{equation}
 \tilde g_1 = \bar g_1 + \left[q_3 e^{-Q} U \right] (x^0_U) \left(\frac{w(x^0_U) - w(x^0_L)}{\bar p_2} - \bar p_3\right)\ .
\end{equation}
Then the expectation values take the form (all quantities are to be taken at the point $x^1$, which is suppressed)
\begin{align}
\label{q1}
\begin{split}
 \langle q_1(x) \rangle &= \tilde g_1 - \frac{\tilde g_3}{\bar p_2}\left([e^QV](x^0_U) - e^QV(x^0)\right) + \frac{\tilde g_3}{\bar p_2} U(x^0) \left(w(x^0_U)-w(x^0)\right) \\
&\quad - [q_3 e^{-Q}](x^0_U) U(x^0)\left(\frac{w(x^0_U) - w(x^0_L)}{\bar p_2} - \bar p_3\right)
\end{split} \\
\label{q2}
\begin{split}
 \langle q_2(x) \rangle &=  \tilde g_2 + \tilde g_1 (x^0_U-x^0)
  - \frac{\tilde g_3}{\bar p_2}\left([e^Q V](x_U^0) (x_U^0 - x^0) - \frac{1}{\bar p_2}\left(w(x^0_U)-w(x^0)\right) \right)\end{split}\\
\label{q3}
 \langle q_3(x) \rangle &= e^Q \tilde g_3(x)\,.
\end{align}

\subsection{Equations of motion}
Restriction of the remaining parameters $\bar p_i$ and $\bar g_i$ in principle should be done via quantum constraints (Ward identities, Schwinger-Dyson equations). However, as we know that the quantum effective action just is the classical action, the $\langle p_i \rangle$ and $\langle q_i \rangle$ for vanishing sources should obey the classical equations of motion,
\begin{align}
\label{eom1}
 \partial_0 q_1 - q_3 \left(\frac{\partial U}{\partial p_1} p_2 p_3 + \frac{\partial V}{\partial p_1} \right) &= 0 \\
\label{eom2}
 \partial_0 q_2 + q_1 - q_3 p_3 U &= 0 \\
\label{eom3}
 \partial_0 q_3 - p_2 q_3 U &= 0 \\
\label{eom4}
 \partial_1 p_1 + p_3 q_3 - p_2 q_2 &= 0  \\
\label{eom5}
 \partial_1 p_2 + q_1 p_2 - q_3 (U p_2 p_3 + V) &= 0 \\
\label{eom6}
 \partial_1 p_3 - q_1 p_3 + q_2 (U p_2 p_3 + V) &= 0\,.
\end{align}
Eq.~\eqref{eom3} is trivially satisfied, \eqref{eom1} and \eqref{eom2} in the special case of $U=0$ are trivial as well. For $U \neq 0$ they both yield the same condition
\begin{equation}
\label{g3tilde2}
 \tilde g_3 = \left[q_3 e^{-Q} \right] (x^0_U)\ .
\end{equation}
Comparison with \eqref{g3tilde1} shows that the condition translates into $\bar g_3 = 0$.
Furthermore \eqref{g3tilde2} allows to simplify \eqref{q1} to
\begin{equation}
 \langle q_1(x) \rangle = \tilde g_1 - \frac{\tilde g_3}{\bar p_2}\left([e^QV](x^0_U) - e^QV\right) + \frac{\tilde g_3}{\bar p_2} U \left(\bar p_2 \bar p_3 - \left(w(x^0) - w(x^0_L) \right)\right)\ .
\end{equation}
There remain the three equations of motion \eqref{eom4}-\eqref{eom6}. As they contain derivatives on the $p_i$ they define the integration functions $\bar g_i$ in terms of $\bar p_i$ and $\partial_1 \bar p_i$. The ensuing relations from \eqref{eom4} and \eqref{eom5} are
\begin{gather}
\label{d1pbar1}
\partial_1 \bar p_1 = \bar p_2 \tilde g_2 + \bar p_2 \tilde g_1 (x^0_U-x^0_L) 
 - \tilde g_3 [e^Q V](x^0_U) (x^0_U-x^0_L) - \frac{\tilde g_3}{\bar p_2} \left(\bar p_2 \bar p_3 - \left(w(x_U^0) - w(x^0_L) \right) \right)\ ,\\
 \label{d2pbar2}
 \partial_1 \bar p_2 = \tilde g_3 [e^Q V](x^0_U) - \bar p_2 \tilde g_1\ . 
\end{gather}
Finally, the combination of \eqref{eom5} and \eqref{eom6} yielding the Casimir function \eqref{casimir},
\begin{equation}
 p_3 \text{\eqref{eom5}} + p_2 \text{\eqref{eom6}} = \partial_1 (p_2 p_3) + \partial_1 p_1 (U p_2 p_3 + V)  = 0\quad
\Rightarrow \quad \partial_1 \left(e^Q p_2 p_3 + \int_{\tilde{X}}^{p_1} dy e^Q V(y) \right) = 0\ ,
\end{equation}
establishes
\begin{equation}
\label{quantC}
 \mathcal C = \bar p_2 \bar p_3 - \left(w(\tilde X) - w(x^0_L) \right)\ .
\end{equation}
The choice of the lower integration limit $\tilde X$ reflects the freedom to shift the Casimir function by a constant.

\subsection{Counting degrees of freedom and comparison with classical solution}
In order to conclude the calculation we should count the number of degrees of freedom that are left after imposing the correct boundary conditions and all restrictions from the equations of motion. Furthermore it should be shown that the obtained solution indeed is equivalent to the classical result.

The mean values \eqref{p1}-\eqref{p3} and \eqref{q1}-\eqref{q3} are parametrized in terms of six $x^1$-dependent functions $\bar p_i$ and $\bar g_i$. At the same time there exist six conditions imposed on these solutions, three boundary conditions and the three differential equations \eqref{eom4}-\eqref{eom6}, resp.\ the two differential equations \eqref{d1pbar1}, \eqref{d2pbar2} and the definition of $\mathcal C$ in \eqref{quantC}.\footnote{Here, we consider the boundary to be a generic boundary rather than a horizon. Then no residual gauge degrees of freedom are left \cite{Bergamin:2005pg}.} Let us first consider the boundary conditions imposed on $p_1$, $q_2$ and $q_3$. They fix uniquely the parameters $\bar p_1$, $\bar g_3$ and $\bar g_2$, 
\begin{align}
\label{cond1}
 p_1(x^0_U) &= \bar p_1 + \bar p_2 (x^0_U-x^0_L) &&\Rightarrow& \bar p_1 &= p_1(x^0_U) - \bar p_2 (x^0_U-x^0_L) \\
\label{cond2}
 q_2(x^0_U) &= \tilde g_2 &&\Rightarrow& \bar g_2 &= 0 \\
\label{cond3}
 q_3(x^0_U) &= e^{Q(x^0_U)} \tilde g_3 && \Rightarrow& \bar g_3 &= 0\,.
\end{align}
Equation \eqref{cond3} is seen to be equivalent to \eqref{g3tilde2}, which followed from the equations of motion alone. We mention that \eqref{cond2} and \eqref{cond3} set two of the three boundary terms in \eqref{newboundary} to zero.

There remain the three differential equations from \eqref{eom4}-\eqref{eom6}. Using equation \eqref{d2pbar2}, $\tilde g_1$ is determined as
\begin{equation}
 \tilde g_1 = \frac{1}{\bar p_2} \left(\tilde g_3 \left[e^Q V\right](x^0_U) - \partial_1 \bar p_2\right)\,,
\end{equation}
and \eqref{d1pbar1}, after inserting the definition \eqref{cond1}, can be used to fix $\bar p_3$ as
\begin{equation}
 \bar p_3 = \frac{1}{\tilde g_3}\left(\bar p_2 \tilde g_2 + \frac{\tilde g_3 \left(w(x^0_U)-w(x^0_L)\right)}{\bar p_2} - \partial_1 p_1(x^0_U)\right)\ .
\end{equation}
As only equation the definition of the Casimir function in \eqref{quantC} is left, which defines $\bar p_2$,
\begin{equation}
 \bar p_2 = \frac{1}{\bar p_3}\left(\mathcal C + w(\tilde X) - w(x^0_L)\right)\,.
\end{equation}
In this equation exactly one constant, namely the value of the Casimir function, can still be chosen, an observation that agrees with the conclusions of \cite{Bergamin:2005pg}.

Finally, we want to compare the obtained solution with the known classical one and identify the different variables. The solution (3.20)-(3.23) of \cite{Grumiller:2002nm} in the gauge \eqref{gauge} is parametrized by one constant, namely the Casimir function, and three $x^1$-dependent functions $\bar X(x^1)$, $X^+(x^1)$ and $F(x^1) = \partial_1 f$, with
\begin{gather}
\begin{alignat}{2}
 X &= \bar X + x^0 X^+\ , &\qquad X^- &= \frac{e^{-Q}}{X^+}(\hat{\mathcal{C}} - \hat w)\ ,
\end{alignat}\\
\begin{alignat}{3}
 \omega_1 &= - \frac{\partial_1 X^+}{X^+} + (U X^+ X^- + V) e^Q F\ , &\qquad e_1^- &= \frac{\partial_1 X}{X^+} + X^- e^Q F\ , &\qquad e_1^+ &= X^+ e^Q F\ .
\end{alignat}
\end{gather}
To distinguish different choices of the integration constants in \eqref{casimir}, hatted symbols were used for the Casimir function $\hat{\mathcal{C}}$ and the invariant potential $\hat w$.
Comparison of this solution with \eqref{p1}-\eqref{p3} and \eqref{q1}-\eqref{q3} yields the following immediate identifications
\begin{align}
\label{cl1}
 \bar p_2 &= X^+\ , & \bar p_1 &= \bar X + X^+ x^0_L\ , & \mathcal C &= \hat{\mathcal{C}} - w(\tilde X)\ , &  w(x^0) &= \hat w(x^0)\ , & \tilde g_3 &= X^+ F\ ,
\end{align}
completed with the conditions for $\tilde g_1$ and $\tilde g_2$:
\begin{gather}
\label{cl2}
\tilde g_1 = - \frac{\partial_1 X^+}{X^+} + \left[e^Q V\right](x^0_U) F \\
\label{cl3}
\tilde g_2 = \frac{1}{X^+}\left(\partial_1 \bar X + \partial_1 X^+ x^0_U + \left(\mathcal C + w(\tilde X) - w(x^0_U)\right)F \right)
\end{gather}
Of course, the quantum result is still parametrized by six $x^1$-dependent functions, but \eqref{cl2} and \eqref{cl3} can easily be rewritten using the identifications \eqref{cl1}, which shows that these conditions are equivalent to \eqref{d2pbar2} and \eqref{d1pbar1}, resp. Together with the condition of $\mathcal C$ being constant full agreement with the classical solution is found, which again confirms the quantum triviality of the model.
\section{Discussion and conclusions}
In this paper we considered the quantization of two-dimensional dilaton gravity with boundaries. The success of our approach relied on a suitable first order formulation of the action and the choice of a spacelike boundary with respect to the evolution parameter of the Hamiltonian analysis. In this case, the whole quantization procedure closely follows earlier results of spacetimes without boundary \cite{Kummer:1997hy,Kummer:1998zs,Grumiller:2002nm}. Still, the ensuing boundary terms of the quantum result are crucial to obtain the correct mean field values, even in the case where the boundary is sent to infinity.

We extended the proof of local quantum triviality to spacetimes with initial and final boundaries (w.r.t. to the Hamiltonian evolution parameter) and showed that the mean fields are compatible with the classical equations of motion. Not surprisingly exactly one boundary degree of freedom, namely the value of the Casimir function, is found to remain on-shell, which agrees with \cite{Bergamin:2005pg} where the same system has been analyzed classically. Still, this result does not yet encompass the full quantum dynamics, as it is valid for vanishing sources, only. In future works a full quantum derivation of all remaining constraints via Ward identities will be important. Also, the consistent treatment of the boundary degrees of freedom in quantum theory should be considered. While it can be freely chosen in the classical theory, an integration over the classical phase space could emerge in quantum theory. Indeed, the path integrations of $q_2$ and $q_3$, which yielded the functional $\delta$-functions \eqref{solp2} and \eqref{solp3}, do not extend to the boundary, as $q_2$ and $q_3$ are held fixed there. Therefore, there should remain a path integral over the boundary values of $p_2$ and $p_3$, eventually leading to a path integration over all values of the Casimir function. Finally, it would be interesting to redo the calculation together with matter fields. In this way, propagating degrees of freedom are added to the theory and the consequences of boundaries on nonperturbative S-matrix elements \cite{Fischer:2001vz,Grumiller:2002nm} could be investigated.

\medskip
{\bf Acknowledgment.} The authors would like to thank D.~Grumiller for numerous discussions and important comments. One of us (L.B.) would like to thank the organizers of GAS@BS 07 for their work making this interesting and stimulating workshop possible.

\end{document}